# TO A PROBLEM OF NOT INCREASING DYNAMIC COMPLIANCE AT ASTHMA PATIENTS WITH VENTILATING DISORDERS AFTER BEROTEC INHALATION.

K. F. Tetenev MD. PhD. Surgut State University. Department of Medicine. Respiratory Division. Surgut. Russia.

Abstract: The purpose of research was to check up the influence of decrease of nonequality of ventilating (after bronchodilator's (berotec) inhalation (BI)) on the magnitude of dynamic compliance of lungs ($C_{dyn}$) at asthma patients with ventilating infringements.
Methods and materials: 20 patients (with 2 and 3 degrees of ventilating infringements (VC<73%, FEV1<51%, MVV<56%), without restrictive disease of lungs, suffering from bronchial asthma were studied before and after BI by plotting volume, rate flow, against the transpulmonare pressure. About the change of nonequality of ventilating we consider by the change after BI of $C_{dyn}$, $C_{dyn}$ at once after flow interruption ($C_{dyn1}$), tissue resistance at inhalation ($R_{ti\ in}$) and exhalation ($R_{ti\ ex}$), parameters of ventilating and general parameters of respiratory mechanics.
Results: the parameters of ventilating were improved (P < 0,05). General parameters of respiratory mechanics also improved. $R_{ti\ in}$ and $R_{ti\ ex}$ are made 0,48±0,16; 1,05±0,25 kPa/l/s before BI and decreased 0,09±0,04; 0,28±0,09 kPa/l/s after BI (P < 0,05; P < 0,05). But $C_{dyn}$ and $C_{dyn1}$ are not changed after BI.
Conclusions: 1. The decrease of ventilation nonequality and tissue friction after BI do not influence on the initially reduced dynamic compliance of lungs at asthma patients without any restrictive diseases of lungs. 2. The cause of not increasing of dynamic compliance after BI probably due by changes in elastic component of parenchyma of lungs, insensitive to berotec.

The decrease of a dynamic compliance ($C_{dyn}$) for the bronchial asthma patients is now established [7, 13]. The decrease of $C_{dyn}$ for these patients is explained by nonequality of ventilation, which is caused by the "pendelluft" phenomenon [5, 7]. The nonequality of ventilating can also be described by increasing of tissue resistance ($R_{ti}$). The increase of ($R_{ti}$) at asthma patients is also established [6, 7]. The decrease of $C_{dyn}$ and the simultaneous increase of $R_{ti}$ were traditionally explained by the development of the ventilation nonequality and friction between good and poorly ventilating parts of lungs (tissue friction) [13].

But the explanations of increase of dynamic elastance with bronchoconstriction often involve presumed stiffening of the lung tissue through interdependence with the airway tree [10]. However Bates et al. concluded that the increases in tissue resistance and elastance (in dogs) are due largely to the development of severe inhomogeneity in the airway tree with progressively isolates parts of the peripheral tissue from the central airways [2]. Kaczka et al. obtained evidence in asthma patients of the existence of widespread peripheral airway constriction leading to shunting of applied flow oscillations to the central airways [5].

Thus the purpose of our research was to test in practice the existence of the cause and effect relationship between the ventilation nonequality of lungs on the one hand, and the decrease of $C_{dyn}$ on the other hand. For this purpose we have decided to investigate the parameters of the respiration mechanics before and after an inhalation of the drug B2 – agonist (berotec). It was supposed, that the improvement of the bronchial conductance should diminish the ventilation nonequality of the lungs, consequently, the $C_{dyn}$ should be increased the $R_{ti}$ should be decreased.

The introduction
Because the lungs compliance is the formula (1) where V – the lung volume (l), P – the transpulmonare pressure (kPa), which one is conditioned by the elastic properties of lungs, tissue friction, and aerodynamic component. The aerodynamic component is unessential in quasistatic conditions (during a slow, deep inspiration).
$C_{dyn} = l/kPa$ (1)
Then:
$C_{dyn} = V/(P_{el} + P_{Rti})$ (2),
Where $P_{el}$ – tpp, which one is necessary for overcoming of elastic properties of lung, $P_{Rti}$ – tpp, conditioned by an expense for overcoming of tissue friction in quasistatic conditions.
For the more exact measurement of a dynamic compliance of lungs we shall enter the magnitude $C_{dyn1}$ – the lungs compliance immediately after an airflow interruption.
The aerodynamic component is unessential in quasistatic conditions; however, it is not equal to zero. The interruption of airflow will allow eliminating an aerodynamic component completely. Thus, the lungs compliance will be influenced by an inhomogeneity of lungs and by the lungs stress-relaxation phenomenon at once after an airflow interruption [8].

There is an improvement of bronchial conductance after berotec inhalation, hence the ventilation nonequality should decrease and the lung compliance should be increased. It is explained by the decrease of the necessary expends for overcoming of tissue friction transpulmonary pressure ($P_{Rti}$) after the inhalation. $C_{dyn}$ is in inverse proportion with tpp; hence, the lungs ventilation will be carried out with the smaller muscles costs. This process is physically described by the formula 2.

On the other hand, the after inhalation bronchial conductance improvement should diminish the "pendelluft" phenomenon essentially, and, hence, reduce the volume leakage. Thus, the measuring lungs volume should be increased. According to the formula 1 $C_{dyn}$ is directly proportional to measuring lungs volume, hence it also should increase. Considering all the above mentioned moments we can conclude, that according to the traditional views there are two reasons for the reduction of the dynamic compliance of the lungs for the asthma patients: the tissue friction increasing and the volume leakage.

$R_{ti}$ – the resistance, which must be surmounted by the respiratory musculation for the overcoming of tissue friction per an airflow unit. The tissue resistance is a part of a nonelastic resistance of the lungs (RL). For its determination it is necessary to measure the RL first. RL is defined by the ratio of the transpulmonare pressure P (kPa) to its appropriate flow V (l/s).

RL=P/V (3)

RL is conditioned by the bronchial resistance (aerodynamic component) ($R_{ac}$) and tissue resistance.

RL=$R_{ac}$+$R_{ti}$ (4)

Then

RL=$P_{ac}$/$V_{ac}$+$P_{ti}$/$V_{ti}$ (5)

Where $P_{ac}$ is tpp, conditioned by an aerodynamic component and volume leakage.
$P_{ti}$ – tpp, caused by tissue friction per an airflow unit $V_{ti}$. $V_{ti}$ also depends on the volume leakage.

However, $V_{ti}$ depends on the volume leakage to a rather lesser extent, than $C_{dyn}$, because the volume leakage also influences an aerodynamic resistance. The aerodynamic resistance is excluded from RL at the measurement of $R_{ti}$.

The aerodynamic resistance is defined during the airflow interruption. The inertial resistance at spontaneous breathing is not essential.

$P_{ac}$ and $V_{ac}$ are excluded at once after an airflow interruption (5), because the aerodynamic resistance is equal to zero after a blocking of the flow by the valve.

Consequently, in these conditions (at once after an airflow interruption) RL=$R_{ti}$=$P_{ti}$/$V_{ti}$ (6).

$R_{ti}$ for the bronchial asthma patients rises as a result of increase of tpp: $P_{ti}$, because the tissue friction increases in the lungs. On the data [9] the tissue friction does not depend on the flow ($V_{ti}$) in the case, if the flow size is not too great. Hence, the increase of the $R_{ti}$ can take place only at the increase of the $P_{ti}$.

The flow is not taken into the calculation of $C_{dyn}$ (2). $C_{dyn}$ was measured under quasistatic conditions during a slow, deep inspiration. $R_{ti}$ – during a spontaneous breathing. It means, that $C_{dyn}$ and $R_{ti}$ are measured under similar conditions. However, the $C_{dyn}$ and $C_{dyn1}$ include the properties of the elastic elements of the lungs tissue ($P_{el}$).

The magnitudes $C_{dyn}$, $C_{dyn1}$, and $R_{ti}$ depend on $P_{ti}$, this fact makes them similar. The $C_{dyn}$ and $C_{dyn1}$ depend on volume leakage in the greater extent, than $R_{ti}$. However, it is possible to assume that $C_{dyn}$ and $C_{dyn1}$ are the more broad conceptions and they include the $R_{ti}$ in their frames. Both magnitudes also depend on bronchial conductance.

So, we can assume that the results of bronchodilatation for asthma patients will be the decrease of ventilation nonequality, the increase of a dynamic compliance (in greater extent) and the decrease of a tissue resistance (in lesser extent).

Clinical material and technique: 12 women and 8 men aged from 38 to 48 (range 59-76 kg) suffering from bronchial asthma and no other respiratory or cardiac diseases were examined in a stable condition. According to spirographic data all the patients had 2 and 3 degrees of ventilating disorders mainly of obstructive type [1]. VC<73%, FEV1<51%, MVV<56% (They are the average magnitudes). All the patients did not have any osteomuscular infringements.

None of the examined patients was receiving oral B2 agonists, theophilline, or systemic steroids prior to the beginning of research. None of the patients has received inhaled short-acting B2 agonist or anticholinergics for 8 hours before the examination and long-acting B2 agonists for 24 hours before the examination. None of the patients had X-ray evidences of pneumofibrosis or others restrictive disease of lungs.

Every patient has given the informed consent for examination.

65 healthy non-smoking people (34 women and 31 men; weight 54-84 kg) aged from 32 to 48 were examined for the control. The main ventilating parameters of lungs VC, FEV1, MVV were measured by spirography before and after (10 minutes) berotec inhalation. The parameters of respiratory mechanics are obtained by plotting of volume against the transpulmonare pressure (tpp) [3, 9, 12] before and after (10 minutes) berotec inhalation. tpp was determined by a difference between esophageal and mouth pressure, using the esophageal balloon. General works of breathing, nonelastic and elastic fractions of work of breathing were calculated from a respiratory loop "pressure-volume". The dynamic compliance of lungs was estimated on the $C_{dyn}$ and $C_{dyn1}$. $C_{dyn}$ was measured in quasistatic conditions during a slow, deep inspiration, at a level of respiratory volume. $C_{dyn1}$ was measured at the moment of an airflow interruption, in the same respiratory maneuver as the $C_{dyn}$ measuring (fig.1).

Fig. 1. Records of volume curve (V) and changes of transpulmonare pressure (P) before and immediately after an airflow interruption, during a slow, deep inspiration.

$C_{dyn}$ was measured before an airflow interruption. $C_{dyn1}$ was measured at the moment of an airflow interruption at the same level of respiratory volume (as at the $C_{dyn}$), for the same patient, in the same respiratory maneuver. Aerodynamic resistance was calculated on tpp and its appropriate lungs volume, at the moment of an airflow interruption.

The $C_{dyn1}$ also was measured in quasystatical conditions, because its magnitude is influenced by inhomogeneity of lungs and by phenomenon of tissue stress-relaxation immediately after an airflow interruption [8]. The airflow interruption eliminates completely in this case just an aerodynamic component of tpp (which can distort the true dynamic compliance of lungs) while measuring $C_{dyn1}$. There fore we consider that $C_{dyn1}$ is more exact characteristic of dynamic compliance rather than $C_{dyn}$.

The common nonelastic resistance (RL) was defined at an inspiration ($RL_{in}$) and at expiration ($Rl_{ex}$). $R_{ti}$ was defined as a difference between the RL and the aerodynamic resistance at an inspiration and expiration of spontaneous respiration [4]. Aerodynamic resistance ($R_{ac}$) was calculating during an airflow interruption on the magnitude of lungs volume and to it conforming aerodynamic component of transpulmonary pressure ($P_{tpp}$) immediately after an airflow interruption.

We used Bodytest to define the bronchial conductance ($R_{aw}$). We used a Fleish tube pneumotahograph with the interrupter of airflow to measure the parameters of respiratory mechanics (Med. phys. instrument. Kazan).

<u>Results</u>: the respiratory discomfort disappeared for 14 patients after berotec inhalation. Though the patients did not feel this respiratory discomfort before berotec inhalation, probably, because they are accustomed to it. The $C_{dyn}$ and $C_{dyn1}$ did not differ from each other. The main ventilation parameters were on the average improved ($P<0.05$) after berotec inhalation, but they have not reached the normal amounts (similar parameters of the control group). $A_g$, $A_n$, $A_{el}$, RL, $R_{aw}$ have changed and varied after berotec inhalation, as it was expected according to the common views on respiration mechanics of the human (table 1).

However, contrary to our expectations, the average values of $C_{dyn}$ and $C_{dyn1}$ have not changed after berotec inhalation (table 2). They also did not differ from each other, further we shall mention the dynamic compliance as $C_{dyn}$. $C_{dyn}$ varied from 0.99 up to 2.88 l/kPa before berotec inhalation and from 0.91 up to 2.55 l/kPa after berotec inhalation. The $R_{aw}$ has decreased for all patients. The $R_{ti}$ decreased on the average, $R_{ti\ in}$ and $R_{ti\ ex}$ were 0.48±0.16; 1.05±0.25 kPa/l/s before berotec inhalation and have made 0.09±0.04; 0.28±0.09 kPa/l/s accordingly after berotec inhalation ($P<0.05$; $P<0.05$)

$R_{ti\ ex}$ varied from 0.055 up to 2.268 kPa/l/s before berotec inhalation. $R_{ti\ ex}$ has decreased after berotec inhalation for all the patients and varied from 0 up to 0.78 kPa/l/s. $R_{ti\ in}$ varied from 0 up to 1.6 kPa/l/s before berotec inhalation and from 0 up to 0.29 kPa/l/s after it. $R_{ti}$ has increased for 2 patients after an inhalation.

<u>Discussion</u>: however, the initially reduced $C_{dyn}$ and $C_{dyn1}$ (the most sensitive parameters for description of ventilating unevenness) have not varied after berotec inhalation.

It would be possible to suspect that the ventilation nonequality has remained at a previous level after berotec inhalation. However, the respiratory discomfort disappeared after berotec inhalation. The main ventilation parameters and $R_{aw}$ have improved; the work of breathing has improved essentially because of a nonelastic fraction of work of breathing. The $R_{ti}$ (in average) has diminished too. All the above-described moments allow making two suppositions:
1. The berotec inhalation diminishes essentially the ventilation nonequality of lungs
2. Hence, the initially decreased of dynamic compliance of lungs at asthma patients, probably does not depend on ventilation nonequality ("pendelluft" phenomena), bronchial conductance and $R_{ti}$.

Then there is the problem: what is the reason for the $C_{dyn}$ to decrease at asthma patients of this group initially? The objective data of previewing points the bronchial asthma patients to the absence of any reasons for irreversible decrease of $C_{dyn}$.

The microscopic research of the pulmonary tissue of asthma patients reveals, that the hyperinflation of lungs intensifies the lung inhomogeneity and results in complication – the rupture of interalveolar septum [7]. These changes in tissue of lungs are irreversible too. But such changes in lung should intensify the ventilation nonequality and also intensify the friction between good and poorly ventilated parts of lungs. This process in lungs should contribute to the decrease of $C_{dyn}$ and to the increase of $R_{ti}$ [13]. In the case of improvement of bronchial conductance after berotec inhalation there should be decrease of the friction between good and poorly ventilated parts of lungs ($R_{ti}$). We also should expect the decrease of gas leakage ("pendelluft" phenomenon), because these two phenomena: frictions between good and poorly ventilated parts of lungs and «pendelluft» phenomenon stipulate one another. According to the formula 2 there should be rather the increase of $C_{dyn}$ just as the decrease of $R_{ti}$ after berotec inhalation.

The $C_{dyn}$ includes the properties of elastic components of lung tissue, according to the formula 2. It probably explains the different behavior of $C_{dyn}$ and $R_{ti}$ after inhalations. The elastic component performs more importance at determining $C_{dyn}$ than $R_{ti}$ (formula 2). Elastic properties of lungs are presented by smooth muscles of bronchial tree and tissue of lungs. Under investigation there had been established that, an isolated contraction of smooth muscles of bronchial tree does not generally have influence on the stiffness of lungs. [7].

Viewing the changes of elastic properties as the functional properties, Smith [11] has come to a conclusion, that the airway constriction below the terminal bronchioles increases elastic resistance of lungs. The gist of this research consisted in a prediction of the increase of lungs elasticity behind from the contraction of the air duct smooth muscles and the blood vessels.

The contractile elements, alveolar duct and the alveolus modify the form of a parenchyma. This is, probably, the reason for the decrease of $C_{dyn}$ for the patients with asthma. However, if we suspect, that the increased tonus of the lung for the asthma patients is a result of the contraction of smooth muscles (contractile elements). There comes a question: why there is no lung stiffness relaxation after the inhalation, just as we observe an improvement of parameters of respiratory mechanics, ventilation, the decrease of $R_{ti}$?

It is possible to suppose that this contraction is probably the results of the irreversible process in lungs, irrelevant with a bronchospasm, but accompanying it.

Berotec is not able to affect on smooth muscles due to probable insensitiveness these muscles.

M. Ludwig [6] demonstrated an important role of parenchyma contractile elements in the rise tissue resistance. The author has come to a conclusion, that the pulmonary tissue (tissue resistance) can influence the respiration mechanics essentially and this influence is not clear yet. She used the capsule technique for assessment tissue resistance.

$R_{ti}$ has increased for 2 patients after an inhalation. The increase of $R_{ti}$ after the improvement of bronchial conductance is difficult to explain from the traditional point of view. And this is especially difficult to explain, because $C_{dyn}$ has not decreased for these patients.

Another causes of such behavior of dynamic compliance are: Due by influence of flow interruption on the magnitudes of $R_{ti}$. And, probably, by different berotec influencing on dynamic compliance and tissue resistance of lungs.

In our previous investigation we have dates about influencing of flow interruption on parameters of respiratory mechanics [12]. But the $C_{dyn}$ and $C_{dyn1}$ did not differ from each other, $C_{dyn1}$ obtained during flow interruption. If the magnitudes $C_{dyn}$ and $C_{dyn\ 1}$ are including elastic properties (formula 2) we can suppose, that $C_{dyn}$ and processes, leading to phenomenon stress relaxation closely connected in both magnitudes. Flow interruption is not substantial in differences behaviors both magnitudes during measurement $C_{dyn}$ and $R_{ti}$.

The influence of berotec inhalation on dynamic compliance is not sufficient due probably elastic properties of parenchyma of lungs. And berotec probably is insensitive to these elastic properties. The cause of this insensitive is not clear.

Conclusions:
1. The decrease of ventilation nonequality and tissue friction after berotec inhalation do not influence on the initially reduced dynamic compliance of lungs at asthma patients without any restrictive diseases of lungs.
2. The cause of not increasing of dynamic compliance after berotec inhalation probably due by changes in elastic component of parenchyma of lungs, insensitive to berotec.

Table 1. Main parameters of ventilating of lungs and mechanics of breathing at asthma patients before (a) and after berotec inhalation (b) (M±m)

| | | Control group (c) | | a | | b | | P(a-b) | P(c-a) | P(c-b) |
|---|---|---|---|---|---|---|---|---|---|---|
| | | N-65 | | n-20 | | n-20 | | | | |

| | | Control | | | | | | | | | |
|---|---|---|---|---|---|---|---|---|---|---|---|
| VC | % | 102,5 | ± | 1,43 | 69,38 | ± | 3,72 | 83,49 | ± | 3,26 | P < 0,05 | P < 0,001 | P < 0,001 |
| FEV1 | % | 102,6 | ± | 1,33 | 45,11 | ± | 5,06 | 58,76 | ± | 6,08 | P < 0,05 | P < 0,001 | P < 0,001 |
| MVV | % | 101,3 | ± | 1,78 | 50,90 | ± | 5,20 | 71,39 | ± | 6,22 | P < 0,05 | P < 0,001 | P < 0,001 |
| Ag | kgm/ min | 0,28 | ± | 0,018 | 1,05 | ± | 0,18 | 0,58 | ± | 0,11 | P < 0,05 | P < 0,001 | P < 0,05 |
| An | kgm/ min | 0,182 | ± | 0,014 | 1,01 | ± | 0,18 | 0,51 | ± | 0,10 | P < 0,05 | P < 0,001 | P < 0,01 |
| Ael | kgm/ min | 0,16 | ± | 0,007 | 0,27 | ± | 0,04 | 0,20 | ± | 0,04 | - | P < 0,05 | - |
| RL in | kPa/l/s | 0,14 | ± | 0,01 | 1,13 | ± | 0,20 | 0,45 | ± | 0,06 | P < 0,01 | P < 0,001 | P < 0,001 |
| RL ex | kPa/l/s | 0,21 | ± | 0,01 | 2,01 | ± | 0,24 | 0,81 | ± | 0,11 | P < 0,001 | P < 0,001 | P < 0,001 |
| $R_{aw}$ | kPa/l/s | 0,146 | ± | 0,018 | 0,615 | ± | 0,085 | 0,374 | ± | 0,029 | P < 0,05 | P < 0,001 | P < 0,001 |

Table 2. The parameters of dynamic compliance of lungs and issue resistance at asthma patients before (a) and after berotec inhalation (b) (M±m)

| | | Control group (c) N-65 | | | a n-20 | | | b n-20 | | | P(a-b) | P(c-a) | P(c-b) |
|---|---|---|---|---|---|---|---|---|---|---|---|---|---|
| Cdyn | l/kPa | 2,682 | ± | 0,092 | 1,81 | ± | 0,23 | 1,83 | ± | 0,18 | - | P < 0,01 | P < 0,001 |
| Cdyn1 | l/kPa | 2,631 | ± | 0,112 | 1,88 | ± | 0,20 | 1,71 | ± | 0,22 | - | P < 0,01 | P < 0,01 |
| Rt in | kPa/l/s | 0,016 | ± | 0,003 | 0,48 | ± | 0,16 | 0,09 | ± | 0,04 | P < 0,05 | P < 0,05 | - |
| Rt ex | kPa/l/s | 0,018 | ± | 0,002 | 1,05 | ± | 0,25 | 0,28 | ± | 0,09 | P < 0,05 | P < 0,001 | P < 0,05 |